\documentclass[12pt]{article}

\newcommand{\tttilde}{\raisebox{-0.25\baselineskip}{\char'176}}

\def\authorinfo#1{\gdef\@authorinfo{#1}}
\authorinfo{}     

\def\keywords#1{
\par\vspace{0.5ex}{\noindent\normalsize\bf Keywords:} #1
\vspace{0.5ex}   
}

\def\acknowledgments{\section*{Acknowledgments}}

\title{The Sound Manifesto\\ \emph{July 2000}}

\author{Michael J.\ O'Donnell and Ilia Bisnovatyi\\[12pt]
        The University of Chicago\\
        Chicago, IL, USA}

\authorinfo{\texttt{michael\_odonnell@acm.org},
            \texttt{http://www.cs.uchicago.edu/\tttilde odonnell/}}

\begin{document}

\maketitle

\begin{abstract}
Computing practice today depends on visual output to drive almost all
user interaction. Other senses, such as audition, may be totally
neglected, or used tangentially, or used in highly restricted
specialized ways. We have excellent audio rendering through D-A
conversion, but we lack rich general facilities for modeling and
manipulating sound comparable in quality and flexibility to graphics. We
need co-ordinated research in several disciplines to improve the use of
sound as an interactive information channel.

Incremental and separate improvements in synthesis, analysis, speech
processing, audiology, acoustics, music, etc. will not alone produce the
radical progress that we seek in sonic practice. We also need to create
a new central topic of study in digital audio research. The new topic
will assimilate the contributions of different disciplines on a common
foundation. The key central concept that we lack is sound as a
general-purpose information channel. We must investigate the structure
of this information channel, which is driven by the co-operative
development of auditory perception and physical sound
production. Particular audible encodings, such as speech and music,
illuminate sonic information by example, but they are no more sufficient
for a characterization than typography is sufficient for a
characterization of visual information.

To develop this new conceptual topic of sonic information structure,
we need to integrate insights from a number of different disciplines
that deal with sound. In particular, we need to co-ordinate central
and foundational studies of the representational models of sound with
specific applications that illuminate the good and bad qualities of
these models. Each natural or artificial process that generates
informative sound, and each perceptual mechanism that derives
information from sound, will teach us something about the right
structure to attribute to the sound itself. The new Sound topic will
combine the work of computer scientists with that of numerical
mathematicians studying sonification, psychologists, linguists,
bioacousticians, and musicians to illuminate the structure of sound
from different angles. Each of these disciplines deals with the use of
sound to carry a different sort of information, under different
requirements and constraints. By combining their insights, we can
learn to understand of the structure of sound in general.
\end{abstract}

\keywords{sound, user interface, audio, signal processing, acoustics}

\section{Sound as a Topic In Computer Science}

A new topic of scholarly research is emerging in computer
science---\emph{sound}. Sound is normally studied from three different
perspectives:
\begin{enumerate}
\item the \emph{production} of sound by vibrating arrays of materials;
\item the \emph{propagation} of sound through the air;
\item the \emph{perception} of sound by the human ear and brain.
\end{enumerate}
The production and propagation of sound are topics in physics, and the
perception of sound is a topic in physiology and psychology. These
three topics are often classified together as \emph{acoustics}, but
they are normally studied in isolation. Essentially, we have studied
three critical things that happen to sound, without studying sound
itself. Computer science provides the right perspective for a new
study of sound as an inherently interesting object, connected
intimately to, but not identified with, the mechanisms of production,
propagation, and perception.

The study of sound is not new, nor is the idea that it should be a topic
in computer science. In 1993, Carla Scaletti and other forward-looking
scholars held a \emph{Workshop on Sound-related
Computation} \cite{SIGSound} where they presented some visionary position
papers announcing the mission of the Association for Computing
Machinery's new SIGSound group. With this Manifesto, we hope to bring the
Sound topic some of the prominent attention that it deserves. This is
the first version of the Manifesto, prepared for presentation at the
conference on \emph{Critical Technologies for the Future of Computing},
part of SPIE's \emph{International Symposium on Optical Science and
Technology} in August 2000. We plan to expand and refine the Manifesto
into a broad mission statement for the research community interested in
sound.

\subsection{Sound as an Information Channel}

Sound is essentially an information channel. This is the hidden reason
why particular phenomena in physics, physiology, and psychology are
classified together as acoustics. The mechanisms for producing and for
perceiving sound have developed co-operatively to allow information to
be encoded in sound. Sound has an inherent information structure,
determined by the combined capabilities of producers and perceivers, in
which various encodings, such as music and speech, are realized. Study
of the information structure of sound properly belongs at the center of
the complex of acoustical topics, integrating insights from the study of
production and of perception, and illuminating them both by explaining
their purpose.

Computer science is accustomed to studying the structure of
information, and algorithmic mechanisms for manipulating that
structure. But the information structure of sound is substantially
different from the channels that computer scientists have studied in
the past. So, computer scientists can contribute a crucial perspective
to the central topic of sound, while sound can stimulate fascinating
new ideas in computer science.

\subsection{Payoff from the Study of Sound}

The systematic study of sound as an information channel will pay off in
at least three important ways.
\begin{enumerate}
\item We will experience the scholarly joy of improving our
  understanding of an important phenomenon.
\item We will generate new techniques of analysis, modeling, and
  synthesis for existing applications of sound---speech recognition
  and synthesis, understanding of animal behavior, creation of music,
  etc.
\item We will exploit sound in human/computer interfaces, augmenting
  the rich interaction already provided by computer graphics, and
  making formerly visual materials accessible to visually handicapped
  people \cite{raman:aui}.
\end{enumerate}

\subsection{Current Specialized Sound Topics}

Scientists already study a number of specialized topics concerning
sound, including speech synthesis and analysis, linguistics,
architectural acoustics, the behavior of musical instruments,
identification of machines from their sounds, analysis of materials from
their sounds, computer music, sonification. All of these specialties
generate insights into the general structure of sound. But computer
music and sonification are particularly important to my proposed
systematic study of sound, because they study sound in general, with
relatively little constraint from the intended application.

Just as typography exercises only a small and specially structured part
of the space of visual symbols, speech exercises only a small and
specially structured part of the space of sounds. On the other hand, the
needs of musical composition and of sonification \cite{icad:report}
demand very broad and general palettes of sound, with the greatest
possible flexibility in exploiting the natural structure of sound.

\section{Overlapping Time Scales in Sound}

There are at least three different time scales for sound, which are
perceived totally differently:
\begin{itemize}
\item In the \emph{sonic} time scale, we sense the frequency of various
sinusoidal components of a sound signal as pitch and color of a
stationary sound. The sonic scale spans about $10^{-4}$ through
$10^{-2}$ seconds. Sequences in the sonic time scale (e.g., sequences of
samples) are not perceived as temporal sequences at all. The sonic time
scale is analogous to the time scale in which the frequency of light is
perceived as affecting color.
\item In the \emph{transient} time scale we perceive changes in the
sinusoidal content of sound as attack, decay, and other transient
effects. The transient time scale spans about $10^{-3}$ through $10^{0}$
seconds. Sequences in the transient time scale are also not perceived as
temporal sequences. The analog in visual perception to the transient
time scale in sound is not obvious, and perhaps there is none.
\item In the \emph{event} time scale, we perceive the relationship
between transients explicitly as sequences in time. The event time scale
spans about $10^{-2}$ through $10^{2}$ seconds. Perceptual time event
sequences tend to combine events perceived by various senses, including
acoustical and visual events.
\end{itemize}
To demonstrate the radically different perceptual structure of the three
different auditory time scales, just play a familiar sound
backwards \cite{ASA:reversal}. Time reversal has no impact on the mix of
frequencies in the sonic time scale. Time-reversed transients change
radically, and it is difficult or impossible to recognize the
correspondence between the forward and reversed versions. Event
sequences simply reverse themselves transparently. For another
experiment, play a sequence of impulses at increasing frequencies. At 5
Hz, we clearly hear a sequence of clicks. At 1000 Hz we clearly hear a
buzzing tone. We cannot locate the transition precisely. For some
interesting portion of the 20--100 Hz region, we can hear a buzzing
tone, and we can also resolve (but not count) individual clicks.

The temporal structure of sound is inherently more complicated than the
temporal structure of vision, and may resemble that of haptic
sensation. Visual perception also involves vibrational and event scales,
but they are separated by orders of magnitude, and may be treated as
independent dimensions. Due to this huge separation, animated graphics
may be framed, and rendered as sequences of static images. No such
framing can satisfy our perception of audible transients. Also,
acoustical phase relationships are reported to the brain, while phase
relationships in light are imperceptible, except to the extent that they
reveal themselves by cancellation before they stimulate the retina.

\section{Descriptive Models of Sound}

At the center of the study of sound lie descriptive models of sound. We
need models that interpolate between the structure of sound producers
and the structure of sound perceivers, capturing the information that is
transferred from one to the other. We already have some excellent models
of sound for certain purposes, but none of them satisfies the need for
general-purpose models at the center.
\begin{enumerate}
\item \emph{Waveform} models of sound as time sequences of amplitude
values (also called \emph{time-domain} models) are ideal for recording
and rendering sound, and for certain mixing and signal-processing
operations. They have little direct relation to either sound production
or perception.
\item \emph{Fourier} models of sound as sums of sinusoidal components at
different frequencies (the \emph{frequency-domain} models) capture some
important qualities of perception at the cochlear (inner-ear) level for
stationary sound. They represent transients and other time-development
as cancellation properties of infinitely extended sinusoids, completely
missing the structure of the perception of sonic change.
\item \emph{Time-frequency} models of sound as sums of time- and
amplitude-modulated sinusoidal components in the form
\begin{displaymath}
\sigma(t)=\sum_p A_p(t)\sin(\Phi_p(t))
\end{displaymath}
In this form, $t$ is \emph{time}; each integer value of $p$ is the index
of a component called a \emph{partial}; $A_p$ and $\Phi_p$ are the
time-varying \emph{amplitude} and \emph{phase} for the $p$th partial;
the derivative $d\Phi_p/dt$ of phase is the time-varying
\emph{frequency} of the partial.
There are infinitely many time-frequency descriptions of the same sound
signal. With appropriate constraints on the $A_p$ and $\Phi_p$,
time-frequency models may be refined to match the cochlear level of
perception very closely, and they represent a very important step in the
refinement of sound models. Time-frequency models fail to capture higher
levels of perceptual organization, particularly the natural relations
between modulators of different partials. Perceptually very simple
sounds have very complex time-frequency descriptions, and some
perceptually natural transformations are very difficult.
\item \emph{Stochastic} \cite{stochastic} models of sound present some or
all of a sound signal as the output of a stochastic process, instead of
giving a numerically precise description. Since sound perception ignores
a lot of the detailed differences between signals with the same
statistical properties, stochastic models can improve the transparency
and representational efficiency of time-frequency models
substantially. The simplest stochastic model presents sound as the sum
of a deterministic component and white noise. More refined models employ
filtered noise to match the spectral profile of a
sound. Lemur \cite{lemur:noise1,lemur:noise2} associates noise with each
partial, by introducing a stochastic component to each phase function
$\Phi_p$. An ideal stochastic model should represent the perceptually
significant qualities of the noise distributions of different partials,
the distinction between random \emph{flutter} in the phase and random
\emph{jitter} in the amplitude, and finally the correlation vs.\
independence between different random components \cite{noise:correlation}.
\item \emph{Excitation/filter} models of sound decompose sound into an
excitation (which might be presented in a time-frequency description)
and a filter with broad frequency response. Excitation/filter models
make it easy to transform a sound by moving the excitation frequency,
while holding the filter resonances constant. This corresponds fairly
closely to the behavior of many instruments and the sung vowels, and
sound perception seems to have adapted itself to such transformations.
\item \emph{Physical} \cite{McIntyre83,Rodet93} models of sound describe the
configurations of vibrating materials that produce sound, usually with
finite-element simulations. Physical models are very useful for
understanding the behavior of real acoustical
instruments \cite{fletcher:physics}, and take an important step in
establishing constrained relationships between the time development of
different sinusoidal partials. But they fail to capture the perceptual
structure of those relationships. It is generally quite difficult to
manipulate a finite-element model in order to accomplish an intuitively
understood change in the sound that it generates. Manipulation of
finite-element physical models is essentially a mathematical analog to
the manufacture of acoustical musical instruments---not an intuitive
task. Waveguide \cite{smith:waveguide} models organize finite-element systems
and improve their computational complexity, but do little to improve
intuitive transparency.
\item \emph{Modal} \cite{Adrien91,ilia:modal} models of sound are based on physical
insights, but they organize sound producers according to their modes of
vibration, rather than their geometrical material structure. Modal
models can use matrices of interactions between different modes to
constrain the structure of transients. This is an excellent idea,
capturing just the sort of interpolation between production and
perception that we need. The details require a lot more study.
\end{enumerate}

\section{Synthesizing Sound}

Each descriptive model leads fairly naturally to synthesis
techniques. While it is very challenging to implement these synthesis
techniques efficiently enough for real-time processing, once the right
models have been identified, concentrated research plus Moore's law will
probably provide us with sufficiently powerful real-time
computations. So, we do not expect the synthesis problem to be a
strategic determinant of the quality of sound processing.

\section{Analyzing Sound}\label{sec:analyzing-sound}

Every descriptive model provides a target for analysis, but the analysis
problems are more deeply challenging than the synthesis problems.  The
analysis problems for time-frequency models of sound, and all of the
more sophisticated models, are extremely underconstrained---each signal
in the time domain has infinitely many descriptions, with many degrees
of freedom. The core of the analysis problem is to find the right
additional constraints.

The current state of the art in sound analysis is based on various types
of windowed Fourier \cite{windowed-fourier} and
wavelet \cite{Graps95,StrangNguyen96} transforms. These methods are all
equivalent to running the input signal through a bank of filters tuned
to various frequencies. All of these methods produce sequences of
quadruples of time, frequency, amplitude and phase. All of them suffer
from the tradeoff between time and frequency resolution inherent in
linear transforms. Given the time-frequency to amplitude-phase function,
several techniques link these points into frequency- and
amplitude-modulated partials, by linking amplitude peaks at nearby
times. McAulay/Quatieri analysis \cite{mcaulay-quatieri} represents the
state of the art in tracing spectral peaks. Marchand has improved the
location of peaks using signal derivatives \cite{marchand}. Heterodyne
analysis refines the two-step process by tracing each partial with a
dynamically tunable filter. Filter-bank techniques can be refined to
approximate cochlear stimulation very accurately, but we may wish to do
better.

It is natural to suppose that, since auditory perception depends only on
information derived from cochlear stimulation, analysis techniques based
on wavelet transforms can provide satisfactory insight into the
structure of sound. On the contrary, we expect that there are
perceptually important structural properties of a sound that do not
affect our perception of that sound, but do affect its perceived
relations to other sounds (e.g., after mixing, reverberation,
distortion, or other naturally occurring transformations). And, there
may be nonlinear analysis techniques that avoid the time-frequency
resolution limits of linear transforms. Notice that current analysis
methods are linear up to the time-frequency to amplitude-phase function,
but the peak tracing methods are nonlinear. We should seek more
profoundly nonlinear methods.

Keep in mind that the problem is not accuracy in reconstructing the
original signal---perfect accuracy in reconstruction is achievable by
filter-bank techniques. The problem is choosing one of the infinitely
many solutions to a highly underconstrained analysis problem. We have no
a priori definition of the ``correct'' analysis, but we might identify
principles that lead to plausible consensus. The best way to seek these
principles is probably through the idea of minimal
description \cite{minimal-description}, but that is only a motivating
idea, and does not prescribe the solution. Roughly, we would like the
analysis corresponding to the simplest quasiphysical (probably modal)
method for generating a sound. The analysis technique postulated here is
almost sure to be ill conditioned (i.e., small changes in the signal
lead to large changes in the analysis) at certain points, particularly
at perceptually ambiguous points. For example, the distinction between a
pair of beating partials and a single amplitude modulated partial will
probably be ill conditioned.

\section{Auditory User Interfaces}

Graphical user interfaces (GUIs) have improved the quality of
computer/human interaction immensely over teletype terminals. Even very
primitive graphics is very effective in displaying the organization of
different logical segments of interaction through the layout of windows
on a display. Sound has the potential to carry similar amounts of
information, but for the most part our computers just beep at
us. Focused computer science research in sound will end this waste of a
sense, with serious auditory user
interfaces \cite{raman:aui,RamanGries95} (AUIs).

Graphical presentations of information enjoy one large structural
advantage over audible presentations: built-in browsability. A static
two-dimensional display is a small database, and a viewer can browse it
easily by moving her head, eyes, and attention. Cartesian coordinates
provide a built-in indexing system. So the overwhelming majority of GUI
interaction involves sequences of static displays. Nature provides a
rich interactive structure even while the display is static.

Sound can only encode information temporally, so AUI software must
provide all browsing and other interactive function
algorithmically. This is both a curse, because it has delayed the
invention of useful general-purpose AUIs, and a blessing, because it
will force us to a deeper understanding of cognitive issues in the
interactive presentation of information. Direct translation from visual
to audible presentation is very valuable \cite{meijer:image-sound},
particularly when the information involved is essentially geometric
(e.g., the location of objects around a person). But the full value of
the AUI requires methods to present abstract information directly to the
ear, without translating from a visual presentation. Anyone who
encounters a descriptive diagram during a session of recording for the
blind experiences the insufficiency of translating information to sound
through a graphical intermediary.

Projects, such as \emph{AsTeR} and \emph{Emacspeak}, have already
demonstrated the feasibility of browsing in an AUI, and the importance
of direct audible presentation with no graphical intermediary. But the
AUI design space has only been sampled, and an immense amount of
research remains before we sense its structure and boundaries.

\section{Linearity vs.\ Nonlinearity in Sound Production}

Sound as an information channel gets much of its structure from the
interaction between linear and nonlinear phenomena. The near-linearity
of many resonant devices (violin strings, air cavities) and of cochlear
receptors makes modulated sinusoids the right structural building blocks
for sound at the lowest level. Modest nonlinearities in sound
production, transmission, and cochlear stimulation produce harmonic
distortion, making harmonic and near-harmonic sequences of sinusoidal
partials crucial organizational units at the next level of
abstraction. Severe nonlinearities in acoustic generators (bow-string,
air-vocal-cord interactions) tend to generate harmonic sequences of
partials. They also produce the attack-transient patterns that are major
carriers of sonic information. We have very little practical
understanding of the natural constraints on the structure of attacks and
other transients, and this lack is largely due to the essential
nonlinearity of practical acoustic generators.

For the purposes of this manifesto, we will focus on linear vs.\
nonlinear theories of acoustic filters. The word ``filter'' suggests
electronic equalizers, noise suppressors, and similar systems. But the
same basic theories apply to all resonating systems, and we are mainly
interested in understanding natural acoustic generators through the
theory of ``filters.''

The theory of linear stationary (i.e., unclocked)
filters \cite{linear-stationary} is a beautiful area of essentially
complete mathematics, based on the LaPlace
transform \cite{laplace-transform}. This theory provides a thorough
understanding of the structure and behavior of all linear resonant
devices, including characterizations of crucial qualities, such as
stability. We have no similarly comprehensive and tractable theory for
nonlinear filters, and there probably is none.

Most of our understanding of nonlinear filters consists of detailed
treatments of special cases, such as the acoustic generators in musical
instruments and vocal tracts. The only completely general theories of
nonlinear filters that we can find are the Volterra and Wiener
theories \cite{volterra-wiener1,volterra-wiener2}, which generalize the
Taylor series approach to nonlinear functions. These theories fail to
provide the precise sorts of characterizations that we depend on in
linear LaPlace theory, and only a small minority of acoustical research
projects employ them.

For research in sound, we need a theory that rivals the perfection of
LaPlace theory, covering some useful subclass of all nonlinear resonant
devices. We do not know precisely how to define this subclass, but it
should include models of typical nonlinear phenemona that are important
to sound, such as nonlinear acoustical generators and saturation effects
in resonators. Volterra and Wiener theories deserve a careful study, in
hopes that they can enlighten the search for a good definition of
nonlinear filters for models of sound. While aiming for a similar degree
of success with LaPlace theory, we should not insist on detailed
analogies between the results. For example, we may find great value in
nonlinear transforms that are not perfectly invertible. And, instead of
characterizing stability, we may decide to restrict attention to a class
of nonlinear filters that are all stable.

\section{Building a Research Community in Sound}

In order to stimulate new insights into the central topic of sound
structure, we need to combine those who are already doing excellent
research in connected areas into a community of sound research. One tool
for forging such a community is a forum for publication. We have
established such a forum for unrefereed reports in the \emph{Computing
Research Repository} (\emph{CoRR}) supported by \emph{ACM}, the
\emph{Los Alamos e-Print archive}, and the \emph{Networked Computer
Science Technical Reference Library} (\emph{NCSTRL}). We encourage
interested authors to contribute to the repository---description and
instructions are found at
\texttt{http://xxx.lanl.gov/archive/cs/intro.html}. We encourage editors
to solicit authors of posted reports to submit to their journals, and we
are looking for resources to found a new journal on sound.

We also encourage computer science research institutions and associations
to recognize sound as a topic similar in scope to computer graphics. In
particular, we encourage ACM to upgrade SIGSound to the full status of
other special interest groups, with formal membership.

\section{Auditory Scene Analysis}

Auditory scene
analysis \cite{Bregmann90,Bregmann93,BrownCooke94,Cooke92,Ellis96I}.
takes a sound signal with one or more channels and computes a
representation of the signal as the sum of sound sources of various
types. Scene analysis is the ultimate sound modeling problem, analogous
to model construction in computer vision, and it will force us to
clarify the structure of general-purpose models of sound. Auditory scene
analysis may try to mimic human perception, or it may aim at other
representations. Scene analysis requires a method for modeling
individual sound sources, and multichannel versions of the problem may
require tracking those sources through space. Source identification and
separation are key subproblems.  In spite of recent increase in
attention, practical results are very limited. Human listeners are far
better at these tasks than any existing software.

In current implementations, the initial stage for extraction is
usually one of the filter-bank analysis methods mentioned above, which
yields a time-frequency analysis. From this representation auditory
features are extracted and grouped into events by exploiting various
cues, such as synchronized energy onset, continuity, common
modulation, harmonic frequency relations, and so
on \cite{Mellinger91,Cooke92}. In addition to filter-bank methods, other
techniques have been employed, such as correlograms and frequency
warping \cite{Wang94}.

The attribution of events to sources is accomplished by providing a set
of rules, which essentially amount to assumptions about the types of
sources allowed in the representation. Variations on the problem deal
with \emph{constrained} or \emph{predetermined sources} that produce a
range of known sounds, \emph{modeled sources}, \emph{spatially
localized sources}, etc. Variant problems may also be characterized by
specific audio \emph{contexts}, such as music, speech, or a particular
environment (traffic, industrial noise, etc.) We may also constrain the
number of sources involved, or we may be interested in tracking only one
or few important sources (e.g. signal-from-noise separation and
voice extraction). The difficulty and scope of the problem solved
and the success of the solution depend greatly on the source
constraints.

It is clear that bottom-up signal processing alone will not yield a
satisfying solution to the problem of auditory scene analysis. Human
listeners are able to employ a variety of higher-level organizational
information in order to facilitate source separation and
identification. Expectations of source behavior are very important and
are affected by previous exposure, aesthetics, training,
etc. Furthermore, different auditory contexts require different amounts
of organizational knowledge and specialized learning: almost anyone can
separate speech from traffic noise, while following the lines played by
different instruments of a string quartet generally requires ear
training and possibly the study of composition. Additional complications
arise from auditory illusions, such as our tendency to infer continuity
and to hear sounds not present in the
source \cite{Bregmann90,asa:masking}.

\section{Aesthetics and Efficacy}

Auditory user interfaces and sonification of scientific data call for
an unusual partnership between science/engineering and the arts, in
which the arts contribute to goals in science and engineering.
Partnerships between scientists and artists are nothing new. The great
computer music institutes, such as IRCAM, CCRMA, \dots have always
fostered such collaboration. Most collaboration aims at using
techniques from science and engineering to understand art, or to
produce art. Problems in musical perception also provide motivation
and paradigmatic examples that drive some scientific research in
auditory perception. The reverse---application of artistic principles
to science and engineering---is relatively rare. We hope to see an
increase in the application of art to science and engineering, based
on the intuitive connection between aesthetics and efficacy in
presenting information.

Scientists and engineers often suffer from a sort of reverse snobbery
when they present their ideas. Overly polished presentations are
sometimes scorned on the assumption that time and effort were applied to
style at the expense of substance. Foolish attempts at cute style can
obscure information, as Tufte shows for graphical
presentations \cite{tufte1,tufte2,tufte3}. On the other hand, crudeness
in presentation also obscures conceptual content. Many early efforts at
visualization of scientific data through 3-dimensional animations were
ineffective due to distractingly poor aesthetics in the production, and
the need for some level of artistic production quality is well accepted
in visualization. We expect a similar need for good aesthetic design
even in the most pragmatically motivated projects in auditory
presentation of information. Because of our poorer understanding of the
structure of sound, compared to visual scenes, We expect an even greater
need for help from artists who enjoy an intuitive ability to construct
aesthetically pleasing sounds. There are already a few collaborations
where musical artists contribute their aesthetic intuitions to
scientific sound projects, and we hope to see more.

So far, we know of no general scientific study of the link between
aesthetics and efficacy. That will be a very exciting new area of
research for anyone who discovers how to approach the issue. There has
already been some study of the value of musical sounds for encoding
nonmusical information \cite{musical-communication}.

\section{World Wide Web References}

\begin{itemize}
\item {ACM SIGSound}
  \begin{itemize}
  \item \texttt{http://www.acm.org/sigsound/}
  \end{itemize}
\item \emph{Computer Music Journal}
  \begin{itemize}
  \item \texttt{http://mitpress.mit.edu/e-journals/Computer-Music-Journal/}
  \end{itemize}
\item \emph{Journal of New Music Research}
  \begin{itemize}
  \item \texttt{http://www.swets.nl/jnmr/jnmr.html}
  \end{itemize}
\item Computer Music Reference Lists
  \begin{itemize}
  \item \texttt{http://mitpress.mit.edu/e-journals/Computer-Music-Journal/}
        \texttt{References/ReferenceList}
  \item \texttt{ftp://ftp.cs.ruu.nl/pub/MIDI/DOC/bibliography.html}
  \end{itemize}
\item International Computer Music Association
  \begin{itemize}
  \item \texttt{http://raven.dartmouth.edu/\tttilde icma/}
  \end{itemize}
\item International Community for Auditory Display
  \begin{itemize}
  \item \texttt{http://www.icad.org/}
  \end{itemize}
\item CoRR archive of Sound research
  \begin{itemize}
  \item \texttt{http://xxx.lanl.gov/archive/cs/intro.html}
  \end{itemize}
\end{itemize}

\acknowledgments

We refined the ideas in \emph{The Sound Manifesto} through discussions
and correspondence with
Larry Fritts,
Josef Jurek,
Hans Kaper,
Gregory Kramer,
Karen Landahl,
Sylvain Marchand,
Daniel Margoliash,
Peter Meijer,
Partha Niyogi,
Howard Nusbaum,
Silvia Pfeiffer,
Stephen Travis Pope,
T. V. Raman,
Davide Rocchesso,
Howard Sandroff,
Carla Scaletti,
Fred Stafford,
Robert Strandh,
Sever Tipei,
Damien Tromeur-Dervout,
Paul Vickers.
Ilia Bisnovatyi prepared most of the sound and graphics demonstrations.

\nocite{*}

\bibliographystyle{plain}

\bibliography{manifesto}

\end{document}